\documentclass{article}
\usepackage{spconf,amsmath,graphicx}
\usepackage[outdir=./]{epstopdf}
\usepackage{bm}
\usepackage{amsfonts}
\usepackage{amssymb}
\usepackage{color}
\usepackage{flushend}
\usepackage{algorithm}
\usepackage{algorithmic}

\title{Hybrid Beamforming With Sub-arrayed MIMO Radar: Enabling Joint Sensing and Communication at mmWave Band}
\name{Fan Liu, Christos Masouros}
\address{Department of Electronic and Electrical Engineering, University College London}
\begin{document}
%
\maketitle
\begin{abstract}
In this paper, we propose a beamforming design for dual-functional radar-communication (DFRC) systems at the millimeter wave (mmWave) band, where hybrid beamforming and sub-arrayed MIMO radar techniques are jointly exploited. We assume that a base station (BS) is serving a user equipment (UE) located in a Non-Line-of-Sight (NLoS) channel, which in the meantime actively detects multiple targets located in a Line-of-Sight (LoS) channel. Given the optimal communication beamformer and the desired radar beampattern, we propose to design the analog and digital beamformers under non-convex constant-modulus (CM) and power constraints, such that the weighted summation of the communication and radar beamforming errors is minimized. The formulated optimization problem can be decomposed into three subproblems, and is solved by the alternating minimization approach. Numerical simulations verify the feasibility of the proposed beamforming design, and show that our approach offers a favorable performance tradeoff between sensing and communication.
\end{abstract}
\section{Introduction}
\label{sec:intro}
To address the explosive growth of wireless devices and services, the coming 5G network aims at a 1000X increase in the capacity by exploiting the large bandwidth available at mmWave band \cite{6515173}. In the meantime, it is expected that the mmWave BS could be equipped with the sensing ability, which may find its usage in a variety of scenarios such as vehicle-to-everything (V2X) communications \cite{7786130,8246850}. In light of the above, it is favourable to have joint radar and communication functionalities deployed on a single hardware platform, which can support simultaneous target detection and downlink communications.
\\\indent Existing contributions for dual-functional radar-communi\\-cation (DFRC) system mainly focus on the applications in the lower frequency bands \cite{mani_dual,opt_waveform_fan,hassanien2016dual}, e.g. sub-6GHz, and are thus difficult to be extended to the mmWave scenarios. Recent works propose to implement the radar function to support V2X communications based on the IEEE 802.11ad WLAN protocol, which operates at the 60GHz band \cite{8114253,8309274}. As the WLAN standard is typically indoor based and employs small-scale antenna arrays, it can only support short-range sensing at the order of tens of meters.
\\\indent To overcome the aforementioned drawbacks, we propose to exploit the large-scale antenna array deployed at the mmWave BS, which can compensate the high path-loss imposed on the mmWave signals. Moreover, the high degrees of freedom (DoFs) of massive antennas make it viable to support joint sensing and communication tasks. In order to reduce the hardware complexity and the associated costs, the hybrid analog-digital (HAD) beamforming structure is typically used in such systems, which requires much less RF chains compared to fully digital tranceivers \cite{8030501}. It is interesting to note that, the HAD system is similar to an existing kind of radar called sub-arrayed MIMO radar that trades-off between phased-array and MIMO radars \cite{6104178}. Therefore, it is natural to combine these two techniques in the design of the mmWave DFRC systems.
\\\indent In this paper, we consider a mmWave DFRC scenario where the BS detects targets while serving a multi-antenna UE in the downlink. We design analog and digital beamformers that can approach a given fully digital communication beamformer, while formulating a desired radar spatial beampattern that points to the target directions. The proposed approach is modeled as a weighted minimization problem, which can be decomposed into three sub-problems. We then solve the problem via a triple alternating minimization method. Simulation results verify the feasibility of the proposed scheme, which achieves a favourable performance tradeoff for radar and communication.
\section{SYSTEM MODEL}
\label{sec:formulation}
We consider a mmWave downlink where an $N_t$-antenna BS serves an $N_r$-antenna UE in the NLoS channel. In the meantime, the BS senses the nearby environment by steering the probing beams to the targets in the LoS channel. An HAD beamforming structure is deployed on the BS, where the number of the RF chains is $N_{RF}$. Without loss of generality, we assume that both the BS and the UE are equipped with uniform linear array (ULA).
\subsection{Communication Model}
The received signal model at the UE can be expressed as
\begin{equation}\label{eq1}
{\mathbf{y}} = \sqrt \rho  {\mathbf{H}}{{\mathbf{F}}_{RF}}{{\mathbf{F}}_{BB}}{\mathbf{s}} + {\mathbf{n}},
\end{equation}
where $\rho$ denotes the average received power, $\mathbf{H} \in \mathbb{C}^{ \times N}$ is the downlink channel matrix, ${\mathbf{F}}_{RF} \in \mathbb{C}^{N \times N_{RF}}$ and ${{\mathbf{F}}_{BB}} \in \mathbb{C}^{N_{RF} \times N_{S}}$ stand for the analog and baseband (digital) beamformers with $N_S$ being the number of data streams, $\mathbf{s} \in \mathbb{C}^{{N_S} \times 1}$ is the transmitted symbol vector, which satisfies that $\mathbb{E}\left( {{\mathbf{s}}{{\mathbf{s}}^H}} \right) = {{\mathbf{I}}_{{N_S}}}$, and finally ${\mathbf{n}} \sim \mathcal{C}\mathcal{N}\left( {{\mathbf{0}},{N_0}{{\mathbf{I}}_N}} \right)$ denotes the Gaussian noise.
\\\indent Following the extended Saleh-Valenzuela model \cite{8052157,7888145}, the narrowband mmWave channel matrix can be expressed as
\begin{equation}\label{eq2}
{\mathbf{H}} = \sqrt {\frac{{{N_t}{N_r}}}{L}} \sum\limits_{l = 1}^L {{\alpha _l}{{\mathbf{a}}_r}\left( {{\theta _{r,l}}} \right){\mathbf{a}}_t^H\left( {{\theta _{t,l}}} \right)} ,
\end{equation}
where $L$ represents the number of scattering paths, ${\alpha _l}$ is the complex gain of the \emph{l}-th path, and ${{\mathbf{a}}_r}\left( {{\theta _{r,l}}} \right)$ and ${\mathbf{a}}_t^H\left( {{\theta _{t,l}}} \right)$ denote the receive and transmit array response vectors, respectively, where $\theta _{r,l}$ and $\theta _{t,l}$ are the azimuth angle of arrival (AoA) and angle of departure (AoD) for the \emph{l}-th path. For an \emph{N}-antenna ULA, the array response vector can be given as
\begin{equation}\label{eq3}
{\mathbf{a}}\left( \theta  \right) = \frac{1}{{\sqrt N }}{\left[ {1,{e^{j\frac{{2\pi }}{\lambda }d\sin \left( \theta  \right)}},...,{e^{j\frac{{2\pi }}{\lambda }d\left( {N - 1} \right)\sin \left( \theta  \right)}}} \right]^T},
\end{equation}
where $d$ and $\lambda$ denote the antenna spacing and the signal wavelength, respectively. Without loss of generality, we set $d = \lambda/2$, and assume that the channel is perfectly known to the BS.
\\\indent While there are several connected patterns between RF chains and antennas, here we consider the partially-connected structure for simplicity, which is also known as the sub-arrayed structure. Each RF chain is connected to $N_t/N_{RF}$ antennas via $N_t/N_{RF}$ phase shifters, which formulate a sub-array. The associated analog beamformer can be given in the form
\begin{equation}\label{eq4}
{{\mathbf{F}}_{RF}} = \left[ {\begin{array}{*{20}{c}}
  {{{\mathbf{t}}_1}}&{\mathbf{0}}& \cdots &{\mathbf{0}} \\
  {\mathbf{0}}&{{{\mathbf{t}}_2}}&{}&{\mathbf{0}} \\
   \vdots &{}& \ddots & \vdots  \\
  {\mathbf{0}}&{\mathbf{0}}& \cdots &{{{\mathbf{t}}_{{N_{RF}}}}}
\end{array}} \right] \in \mathbb{C}^{N \times N_{RF}},
\end{equation}
where $\mathbf{t}_i \in {\mathbb{C}^{\frac{{{N_t}}}{{{N_{RF}}}} \times 1}}, i = 1,...,N_{RF}$ denotes the values of the phase shifters at the \emph{i}-th sub-array, which contains constant-modulus (CM) entries.
\subsection{Radar Model}
The above partially connected structure corresponds to an existing kind of radar called sub-arrayed MIMO radar. According to \cite{6104178,5419124}, phased-array and MIMO radars can be viewed as the special cases of the sub-arrayed radar by letting $N_{RF} = 1$ and $N_{RF} = N_t$, respectively. Hence, the sub-arrayed radar trades-off between the high directionality of the phased-array radar and the high degrees of freedom (DoFs) of the MIMO radar. The transmit beampattern of the radar can be given as \cite{6104178}
\begin{equation}\label{eq5}
G\left( \theta  \right) = {\mathbf{a}}_t^H\left( \theta  \right){{\mathbf{R}}}{{\mathbf{a}}_t}\left( \theta  \right),
\end{equation}
where ${{\mathbf{R}}} \in \mathbb{C}^{N \times N}$ is the covariance matrix of the precoded waveform, and can be expressed as
\begin{equation}\label{eq6}
\begin{gathered}
  {\mathbf{R}} = \mathbb{E}\left( {{{\mathbf{F}}_{RF}}{{\mathbf{F}}_{BB}}{\mathbf{s}}{{\mathbf{s}}^H}{\mathbf{F}}_{BB}^H{\mathbf{F}}_{RF}^H} \right) \hfill \\
   = {{\mathbf{F}}_{RF}}{{\mathbf{F}}_{BB}}\mathbb{E}\left( {{\mathbf{s}}{{\mathbf{s}}^H}} \right){\mathbf{F}}_{BB}^H{\mathbf{F}}_{RF}^H = {{\mathbf{F}}_{RF}}{{\mathbf{F}}_{BB}}{\mathbf{F}}_{BB}^H{\mathbf{F}}_{RF}^H. \hfill \\
\end{gathered}
\end{equation}
It can be seen from (\ref{eq5}) that to design the radar beampattern is equivalent to designing the covariance matrix above.
\\\indent Suppose that there are $N_{tar}$ targets located at the angles $\left\{ {{\phi _1},{\phi _2},...,{\phi _{{N_{tar}}}}} \right\}$. The typical sub-arrayed MIMO radar beamformer can be formulated as \cite{5419124}
\begin{equation}\label{eq7}
{{\mathbf{F}}_{rad}} = \left[ {\begin{array}{*{20}{c}}
  {{{\mathbf{v}}_1}}&{\mathbf{0}}& \cdots &{\mathbf{0}} \\
  {\mathbf{0}}&{{{\mathbf{v}}_2}}&{}&{\mathbf{0}} \\
   \vdots &{}& \ddots & \vdots  \\
  {\mathbf{0}}&{\mathbf{0}}& \cdots &{{{\mathbf{v}}_{{N_{tar}}}}}
\end{array}} \right] \in \mathbb{C}^{N \times N_{tar}},
\end{equation}
where ${{\mathbf{v}}_i} \in {\mathbb{C}^{\frac{{{N_t}}}{{{N_{tar}}}} \times 1}}$ is composed by the entries of ${{\mathbf{a}}_t}\left( \phi _i \right)$ that are located at the corresponding slots. The associated radar covariance matrix is therefore given by ${{\mathbf{R}}_d} = {{\mathbf{F}}_{rad}}{\mathbf{F}}_{rad}^H$, which is a rank-$N_{tar}$ semidefinite matrix.
\subsection{Problem Formulation}
Our aim is to design the analog and digital beamformers, such that a high-quality communication link can be established between the BS and the UE, while a well-designed radar beampattern is formulated at the BS simultaneously. To guarantee the communication performance, the hybrid beamformer ${\mathbf{F}}_{RF}{\mathbf{F}}_{BB}$ should approach the fully-digital communication beamformer ${\mathbf{F}}_{com}$. Noting the fact that multiplying ${{\mathbf{F}}_{rad}}$ with an unitary matrix $\mathbf{U}$ will not change the resultant radar beampattern, ${\mathbf{F}}_{RF}{\mathbf{F}}_{BB}$ needs to approach ${{\mathbf{F}}_{rad}}\mathbf{U}$ for ensuring the radar performance, where $\mathbf{U} \in \mathbb{C}^{N_{tar} \times N_{S}}$ is an auxiliary unitary matrix variable. We therefore consider the following optimization problem
\begin{equation}\label{eq8}
\begin{gathered}
  \mathop {\min }\limits_{{{\mathbf{F}}_{RF}},{{\mathbf{F}}_{BB}},{\mathbf{U}}} \;\eta \left\| {{{\mathbf{F}}_{RF}}{{\mathbf{F}}_{BB}} - {{\mathbf{F}}_{com}}} \right\|_F^2 \hfill \\
   \;\;\;\;\;\;\;\;\;\;\;\;\;\;\;\; + \left( {1 - \eta } \right)\left\| {{{\mathbf{F}}_{RF}}{{\mathbf{F}}_{BB}} - {{\mathbf{F}}_{rad}}{\mathbf{U}}} \right\|_F^2 \hfill \\
  \;s.t.\;\;\;{{\mathbf{F}}_{RF}} \in {\mathcal{A}_p},\;\left\| {{{\mathbf{F}}_{RF}}{{\mathbf{F}}_{BB}}} \right\|_F^2 = {P_T},{\mathbf{U}}{{\mathbf{U}}^H} = {{\mathbf{I}}_{{N_{tar}}}}, \hfill \\
\end{gathered}
\end{equation}
where $\eta \in \left[0,1\right]$ is a weighting factor that determines the weights for radar and communication performance, ${\mathcal{A}_p}$ represents the feasible set of partially-connected analog beamformers where CM constraints are imposed on the non-zero elements of ${\mathbf{F}}_{RF}$, and finally ${P_T}$ denotes the power budget. Without loss of generality, we assume that $N_{S} \ge N_{tar}$. Note that we enforce an equality constraint for the transmit power, as the radar is often required to transmit at its maximum available power in practice \cite{4276989}.
\\\indent Due to the non-convexities in both the constraints and the objective function, the problem (\ref{eq8}) is rather difficult to tackle. While the global minimizer for (\ref{eq8}) is in general unobtainable, we propose in the following a triple alternating minimization (TAltMin) method that can efficiently yield a near-optimal solution to the problem.
\section{PROPOSED APPROACH}
\label{sec:proposed}
By exploiting the special structure of ${\mathbf{F}}_{RF}$, the power constraint of (\ref{eq8}) can be recast as
\begin{equation}\label{eq9}
\left\| {{{\mathbf{F}}_{RF}}{{\mathbf{F}}_{BB}}} \right\|_F^2 = \frac{{{N_t}}}{{{N_{RF}}}}\left\| {{{\mathbf{F}}_{BB}}} \right\|_F^2 = {P_T}.
\end{equation}
This indicates that the three variables are in fact separable, which yield three sub-problems that are much easier to solve.
\subsection{Sub-problem for $\mathbf{U}$}
By fixing ${\mathbf{F}}_{RF}$ and ${\mathbf{F}}_{BB}$, problem (\ref{eq8}) is equivalent to
\begin{equation}\label{eq10}
\mathop {\min }\limits_{\mathbf{U}} \;\left\| {{{\mathbf{F}}_{rad}}{\mathbf{U}} - {{\mathbf{F}}_{RF}}{{\mathbf{F}}_{BB}}} \right\|_F^2{\text{ }}s.t.\;\;{\mathbf{U}}{{\mathbf{U}}^H} = {{\mathbf{I}}_{{N_{tar}}}}.
\end{equation}
Problem (\ref{eq10}) tends to be a least-squares (LS) problem defined on the Stiefel manifold, which is obviously non-convex. Nevertheless, it has been proven that (\ref{eq10}) can be classified into the so-called Orthogonal Procrustes problem (OPP) \cite{viklands2008algorithms}, which can be optimally solved in closed-form via singular value decomposition (SVD). This is given as
\begin{equation}\label{eq11}
{\mathbf{U}} = {\mathbf{\tilde U}}{{\mathbf{I}}_{{N_{tar}} \times {N_S}}}{\mathbf{\tilde V}},
\end{equation}
where ${\mathbf{\tilde U\Sigma \tilde V}} = {\mathbf{F}}_{rad}^H{{\mathbf{F}}_{RF}}{{\mathbf{F}}_{BB}}$ is the SVD of ${\mathbf{F}}_{rad}^H{{\mathbf{F}}_{RF}}{{\mathbf{F}}_{BB}}$, and ${{\mathbf{I}}_{{N_{tar}} \times {N_S}}}$ is composed by an ${N_{tar}} \times {N_{tar}}$ identity matrix and an ${N_{tar}} \times \left( {{N_S} - {N_{tar}}} \right)$ zero matrix.
\subsection{Sub-problem for ${\mathbf{F}}_{RF}$}
We then fix $\mathbf{U}$ and ${\mathbf{F}}_{BB}$ and solve for ${\mathbf{F}}_{RF}$, in which case the original problem can be reformulated as
\begin{equation}\label{eq12}
\begin{gathered}
  \mathop {\min }\limits_{{{\mathbf{F}}_{RF}}} \;\eta \left\| {{{\mathbf{F}}_{RF}}{{\mathbf{F}}_{BB}} - {{\mathbf{F}}_{com}}} \right\|_F^2\; \hfill \\
  \;\;\;\;\;\; + \left( {1 - \eta } \right)\left\| {{{\mathbf{F}}_{RF}}{{\mathbf{F}}_{BB}} - {{\mathbf{F}}_{rad}}{\mathbf{U}}} \right\|_F^2\;s.t.\;\;\;{{\mathbf{F}}_{RF}} \in {\mathcal{A}_p}, \hfill \\
\end{gathered}
\end{equation}
Again, due to the structure of ${\mathbf{F}}_{RF}$, problem (\ref{eq12}) can be solved in a row-wise manner by solving for each non-zero entry of ${\mathbf{F}}_{RF}$, which is obtained as
\begin{equation}\label{eq13}
\begin{gathered}
  \mathop {\min }\limits_{{\varphi _{i,l}}} \eta \left\| {{e^{j{\varphi _{i,l}}}}{{\left( {{{\mathbf{F}}_{BB}}} \right)}_{l,:}} - {{\left( {{{\mathbf{F}}_{com}}} \right)}_{i,:}}} \right\|_F^2 \hfill \\
  \;\;\;\;\;\;\; + \left( {1 - \eta } \right)\left\| {{e^{j{\varphi _{i,l}}}}{{\left( {{{\mathbf{F}}_{BB}}} \right)}_{l,:}} - {{\left( {{{\mathbf{F}}_{rad}}{\mathbf{U}}} \right)}_{i,:}}} \right\|_F^2, \hfill \\
\end{gathered}
\end{equation}
where $\varphi _{i,l}$ is the phase of the $\left(i,l\right)$ non-zero element of ${\mathbf{F}}_{RF}$, and $\left(\cdot\right)_{i,:}$ denotes the \emph{i}-th row for the matrix. Problem (\ref{eq13}) is nothing but a phase rotation problem, where the optimal solution is given by
\begin{equation}\label{eq14}
{\left( {{{\mathbf{F}}_{RF}}} \right)_{i,l}} = \exp \left( {j\arg \left\{ {{{\mathbf{a}}^H}{\mathbf{b}}} \right\}} \right),
\end{equation}
where
\begin{equation}\label{eq15}
\begin{gathered}
  {\mathbf{a}} = {\left[ {\sqrt \eta  {{\left( {{{\mathbf{F}}_{com}}} \right)}_{i,:}},\sqrt {1 - \eta } {{\left( {{{\mathbf{F}}_{rad}}{\mathbf{U}}} \right)}_{i,:}}} \right]^T}, \hfill \\
  {\mathbf{b}} = {\left[ {\sqrt \eta  {{\left( {{{\mathbf{F}}_{BB}}} \right)}_{l,:}},\sqrt {1 - \eta } {{\left( {{{\mathbf{F}}_{BB}}} \right)}_{l,:}}} \right]^T}. \hfill \\
\end{gathered}
\end{equation}

\subsection{Sub-problem for ${\mathbf{F}}_{BB}$}
Finally, it remains to obtain ${\mathbf{F}}_{BB}$ while $\mathbf{U}$ and ${\mathbf{F}}_{RF}$ are fixed. By recalling (\ref{eq9}), the corresponding sub-problem is
\begin{equation}\label{eq16}
\begin{gathered}
  \mathop {\min }\limits_{{{\mathbf{F}}_{BB}}} \;\eta \left\| {{{\mathbf{F}}_{RF}}{{\mathbf{F}}_{BB}} - {{\mathbf{F}}_{com}}} \right\|_F^2\; \hfill \\
   + \left( {1 - \eta } \right)\left\| {{{\mathbf{F}}_{RF}}{{\mathbf{F}}_{BB}} - {{\mathbf{F}}_{rad}}{\mathbf{U}}} \right\|_F^2\;s.t.\left\| {{{\mathbf{F}}_{BB}}} \right\|_F^2 = \frac{{{N_{RF}}{P_T}}}{{{N_t}}}. \hfill \\
\end{gathered}
\end{equation}
To simplify the problem, let us denote
\begin{equation}\label{eq17}
\begin{gathered}
  {\mathbf{A}} = {\left[ {\sqrt \eta  {\mathbf{F}}_{RF}^T,\sqrt {1 - \eta } {\mathbf{F}}_{RF}^T} \right]^T}, \hfill \\
  {\mathbf{B}} = {\left[ {\sqrt \eta  {\mathbf{F}}_{com}^T,\sqrt {1 - \eta } {{\mathbf{U}}^T}{\mathbf{F}}_{rad}^T} \right]^T}. \hfill \\
\end{gathered}
\end{equation}
Then problem (\ref{eq16}) can be written compactly as
\begin{equation}\label{eq18}
\mathop {\min }\limits_{{{\mathbf{F}}_{BB}}} \left\| {{\mathbf{A}}{{\mathbf{F}}_{BB}} - {\mathbf{B}}} \right\|_F^2\;s.t.\;\left\| {{{\mathbf{F}}_{BB}}} \right\|_F^2 = \frac{{{N_{RF}}{P_T}}}{{{N_t}}},
\end{equation}
which is an LS problem on the complex sphere. By further expanding the objective function, problem (\ref{eq18}) can be rewritten as
\begin{equation}\label{eq19}
\begin{gathered}
  \mathop {\min }\limits_{{{\mathbf{F}}_{BB}}} \;\operatorname{tr} \left( {{\mathbf{F}}_{BB}^H{\mathbf{Q}}{{\mathbf{F}}_{BB}}} \right) - 2\operatorname{Re} \left( {\operatorname{tr} \left( {{\mathbf{F}}_{BB}^H{\mathbf{G}}} \right)} \right) \hfill \\
  s.t.\;\;\left\| {{{\mathbf{F}}_{BB}}} \right\|_F^2 = \frac{{{N_{RF}}{P_T}}}{{{N_t}}}, \hfill \\
\end{gathered}
\end{equation}
where ${\mathbf{Q}} = {{\mathbf{A}}^H}{\mathbf{A}},{\mathbf{G}} = {{\mathbf{A}}^H}{\mathbf{B}}$. Since $\mathbf{Q}$ is a Hermitian matrix, problem (\ref{eq19}) can be regarded as the matrix version of the trust-region sub-problem (TRS), for which strong duality holds \cite{fortin2004trust}. Hence, it is possible to obtain the global minimum of (\ref{eq19}) by solving the associated Karush-Kuhn-Tucker (KKT) equations. Here we adopt [Algorithm 1, 5] to solve the problem, where low-complexity operations such as eigenvalue decomposition and golden-section search are involved in the process. We refer the readers to \cite{opt_waveform_fan} for more details.
\subsection{The TAltMin Procedure}
Now we are ready to describe the TAltMin technique, which has been summarized in Algorithm 1 as follows.
\renewcommand{\algorithmicrequire}{\textbf{Input:}}
\renewcommand{\algorithmicensure}{\textbf{Output:}}
\begin{algorithm}
\caption{Triple Alternating Minimization Algorithm for Solving (\ref{eq8})}
\label{alg:A}
\begin{algorithmic}
    \REQUIRE $\mathbf{H},{\mathbf{F}}_{com},{\mathbf{F}}_{rad}$, $0 \le \eta \le 1$, $P_T$, tolerable accuracy $\varepsilon > 0$ and the maximum iteration number $k_{\max}$
    \ENSURE ${\mathbf{F}}_{RF}$, ${\mathbf{F}}_{BB}$, $\mathbf{U}$
    \STATE 1. Initialize randomly ${\mathbf{U}}^{(0)}$, ${\mathbf{F}}_{RF}^{(0)}$ and ${\mathbf{F}}_{BB}^{(0)}$. Compute the objective function of (\ref{eq8}), denoted as $f^{(0)}$. Set $k=1$.
    \WHILE{$k\le k_{max}$ and $ \left| {{f^{(k)}} - {f^{(k - 1)}}} \right|\ge {\varepsilon}$}
    \STATE 2. Compute ${\mathbf{U}}^{(k)}$ by solving sub-problem (\ref{eq10}).
    \STATE 3. Compute ${\mathbf{F}}_{RF}^{(k)}$ by solving sub-problem (\ref{eq12}).
    \STATE 4. Compute ${\mathbf{F}}_{BB}^{(k)}$ by solving sub-problem (\ref{eq16}).
    \STATE 5. Compute the objective function $f^{(k)}$ based on the obtained variables.
    \STATE 6. $k = k + 1.$
    \ENDWHILE
\end{algorithmic}
\end{algorithm}

The proposed TAltMin algorithm can be viewed as a coordinate descent method, where the convergence can be strictly guaranteed \cite{altmin}. In our simulations, we see that Algorithm 1 always converges within tens of iterations.
\begin{figure}[htp]
 \centering
 \includegraphics[width=0.93\columnwidth]{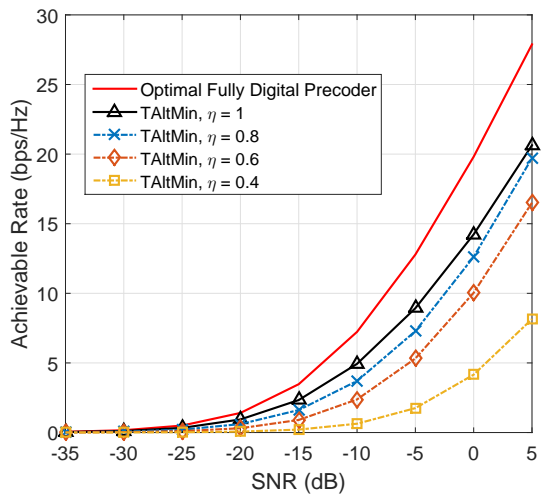}
 \caption{Achievable rate of the mmWave downlink, $N_t = 120$, $N_S = N_r = 6$, $N_{RF} = 24$.}
 \label{fig:FIG2}
\end{figure}
\section{Numerical Results}
\label{sec:results}
In this section, we validate the effectiveness of the proposed beamforming approach via Monte Carlo simulations. Without loss of generality, we assume that $N_t = 120$, $N_S = N_r = 6$, $N_{RF} = 24$, and set $P_T = N_r$ as the normalized power. In our simulations, we consider a mmWave channel with $L = 10$ scattering paths. The number of targets in the LoS channel is set as 3, which are supposed to be located at the angles $\left[-30^\circ, 0^\circ, 30^\circ\right]$. Following the standard assumptions, we assume that each $\alpha_l$ of the mmWave channel subjects to the standard complex Gaussian distribution, and all the AoAs and AoDs follow the uniform distribution on $\left[-\pi,\pi\right]$.
\\\indent We first show the achievable communication rate in Fig. 1 by varying the weighting factor from 0.4 to 1, where the fully digital beamformer at the BS and the combiner at the UE are obtained as the first $N_r$ right and left singular vectors, respectively. It can be observed that with the increase of $\eta$, more weight is allocated to minimizing the Euclidean distance between the optimal and the designed communication beamformers, and hence the achievable rate is on the rise. By letting $\eta = 1$, we obtain the communication-only performance with partially-connected hybrid beamformer, where the radar performance is not addressed. Fig. 2 further illustrates the associated radar beampatterns, where the weighting factor is set as 0.4 and 0.6, respectively. We see that the BS can effectively steer its beams towards the directions of targets while preserving the communication performance. Moreover, by using the proposed TAltMin method, a flexible performance tradeoff can be readily achieved.
\begin{figure}[htp]
 \centering
 \includegraphics[width=0.93\columnwidth]{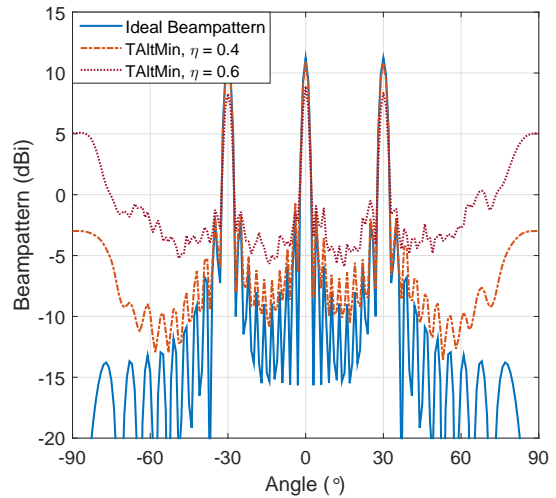}
 \caption{Radar beampatterns obtained by TAltMin, $N_t = 120$, $N_S = N_r = 6$, $N_{RF} = 24$.}
 \label{fig:FIG3}
\end{figure}

\section{Conclusion}
\label{sec:conclusion}
In this paper, we consider the beamforming design for the mmWave downlink, where a hybrid analog-digital beamforming structure is deployed on the BS, which is adopted for accomplishing the joint sensing and communication tasks. While the formulated beamforming optimization is nonconvex, we propose a triple alternating minimization (TAltMin) approach to find a near-optimal solution to the problem. Numerical results show that by using the proposed method, a favorable performance tradeoff can be realized for target detection and downlink communication.



\vfill\pagebreak

%

\bibliographystyle{IEEEbib}
\bibliography{refs}

\begin{thebibliography}{10}

\bibitem{6515173}
T.~S. Rappaport, S.~Sun, R.~Mayzus, H.~Zhao, Y.~Azar, K.~Wang, G.~N. Wong,
  J.~K. Schulz, M.~Samimi, and F.~Gutierrez,
\newblock ``Millimeter wave mobile communications for 5{G} cellular: {I}t will
  work!,''
\newblock {\em IEEE Access}, vol. 1, pp. 335--349, May 2013.

\bibitem{7786130}
J.~Choi, V.~Va, N.~Gonzalez-Prelcic, R.~Daniels, C.~R. Bhat, and R.~W. Heath,
\newblock ``Millimeter-wave vehicular communication to support massive
  automotive sensing,''
\newblock {\em IEEE Commun. Mag.}, vol. 54, no. 12, pp. 160--167, Dec 2016.

\bibitem{8246850}
H.~Wymeersch, G.~Seco-Granados, G.~Destino, D.~Dardari, and F.~Tufvesson,
\newblock ``5{G} mm{W}ave positioning for vehicular networks,''
\newblock {\em IEEE Wireless Commun.}, vol. 24, no. 6, pp. 80--86, Dec 2017.

\bibitem{mani_dual}
F.~Liu, C.~Masouros, A.~Li, H.~Sun, and L.~Hanzo,
\newblock ``{MU}-{MIMO} communications with {MIMO} radar: {F}rom co-existence
  to joint transmission,''
\newblock {\em IEEE Trans. Wireless Commun.}, vol. 17, no. 4, pp. 2755--2770,
  Apr 2018.

\bibitem{opt_waveform_fan}
F.~Liu, L.~Zhou, C.~Masouros, A.~Li, W.~Luo, and A.~Petropulu,
\newblock ``Toward dual-functional radar-communication systems: {O}ptimal
  waveform design,''
\newblock {\em IEEE Trans. Signal Process.}, vol. 66, no. 16, pp. 4264--4279,
  Aug 2018.

\bibitem{hassanien2016dual}
A.~Hassanien, M.~G. Amin, Y.~D. Zhang, and F.~Ahmad,
\newblock ``Dual-function radar-communications: {I}nformation embedding using
  sidelobe control and waveform diversity.,''
\newblock {\em IEEE Trans. Signal Process.}, vol. 64, no. 8, pp. 2168--2181,
  Apr 2016.

\bibitem{8114253}
P.~Kumari, J.~Choi, N.~Gonz¨¢lez-Prelcic, and R.~W. Heath,
\newblock ``{IEEE} 802.11ad-based radar: {A}n approach to joint vehicular
  communication-radar system,''
\newblock {\em IEEE Trans. Veh. Technol.}, vol. 67, no. 4, pp. 3012--3027, Apr
  2018.

\bibitem{8309274}
E.~Grossi, M.~Lops, L.~Venturino, and A.~Zappone,
\newblock ``Opportunistic radar in {IEEE} 802.11ad networks,''
\newblock {\em IEEE Trans. Signal Process.}, vol. 66, no. 9, pp. 2441--2454,
  May 2018.

\bibitem{8030501}
A.~F. Molisch, V.~V. Ratnam, S.~Han, Z.~Li, S.~L.~H. Nguyen, L.~Li, and
  K.~Haneda,
\newblock ``Hybrid beamforming for massive {MIMO}: {A} survey,''
\newblock {\em IEEE Commun. Mag.}, vol. 55, no. 9, pp. 134--141, Sept 2017.

\bibitem{6104178}
D.~Wilcox and M.~Sellathurai,
\newblock ``On {MIMO} radar subarrayed transmit beamforming,''
\newblock {\em IEEE Trans. Signal Process.}, vol. 60, no. 4, pp. 2076--2081,
  Apr 2012.

\bibitem{8052157}
D.~Zhang, Y.~Wang, X.~Li, and W.~Xiang,
\newblock ``Hybridly connected structure for hybrid beamforming in mmwave
  massive {MIMO} systems,''
\newblock {\em IEEE Trans. Commun.}, vol. 66, no. 2, pp. 662--674, Feb 2018.

\bibitem{7888145}
N.~Gonz¨¢lez-Prelcic, R.~M¨¦ndez-Rial, and R.~W. Heath,
\newblock ``Radar aided beam alignment in mm{W}ave {V}2{I} communications
  supporting antenna diversity,''
\newblock in {\em Proc. Information Theory and Applications Workshop (ITA)},
  Jan 2016, pp. 1--7.

\bibitem{5419124}
A.~Hassanien and S.~A. Vorobyov,
\newblock ``Phased-{MIMO} radar: {A} tradeoff between phased-array and {MIMO}
  radars,''
\newblock {\em IEEE Trans. Signal Process.}, vol. 58, no. 6, pp. 3137--3151,
  Jun 2010.

\bibitem{4276989}
P.~Stoica, J.~Li, and Y.~Xie,
\newblock ``On probing signal design for {MIMO} radar,''
\newblock {\em IEEE Trans. Signal Process.}, vol. 55, no. 8, pp. 4151--4161,
  Aug 2007.

\bibitem{viklands2008algorithms}
T.~Viklands,
\newblock {\em Algorithms for the weighted orthogonal {P}rocrustes problem and
  other least squares problems},
\newblock Ph.D. thesis, Comput. Sci. Dept., Umea Univ., Umea, Sweden, 2008.

\bibitem{fortin2004trust}
C.~Fortin and H.~Wolkowicz,
\newblock ``The trust region subproblem and semidefinite programming,''
\newblock {\em Optim. Methods Softw.}, vol. 19, no. 1, pp. 41--67, 2004.

\bibitem{altmin}
A.~Beck,
\newblock ``On the convergence of alternating minimization for convex
  programming with applications to iteratively reweighted least squares and
  decomposition schemes,''
\newblock {\em SIAM J. Optim.}, vol. 25, no. 1, pp. 185--209, 2015.

\end{thebibliography}

\end{document}